\documentclass[useAMS,usegraphicx,usenatbib]{mn2e}

\newcommand{\oii}{\mbox{${\rm [OII]\lambda3727}$}}
\newcommand{\oiiis}{\mbox{${\rm [OIII]\lambda5007}$}}
\newcommand{\oiii}{\mbox{${\rm [OIII]\lambda4959,\lambda5007}$}}

\newcommand{\halpha}{\mbox{${\rm H\alpha}$}}
\newcommand{\hbeta}{\mbox{${\rm H\beta}$}}
\newcommand{\rp}{\mbox{${\rm R_{23}}$}}
\newcommand{\doh}{\mbox{${\rm 12+\log(O/H)}$}}

\newcommand{\oo}{\mbox{${\rm O_{32}}$}}
\newcommand{\vrot}{\mbox{${V_{\rm rot}}$}}
\newcommand{\rdspec}{\mbox{${r_{\rm d,spec}}$}}

\newcommand{\slantfrac}[2]{#1\!\left/#2\right.}
\newcommand{\snr}{$\slantfrac{S}{N}$}

\newcommand{\zcl}{z_{\rmn{cl}}}
\newcommand{\sigmacl}{\sigma_{\rmn{cl}}}
\newcommand{\unisim}{\sim\!}

\title{Star formation rates and chemical abundances of emission line galaxies
  in intermediate-redshift clusters}

\author[Mouhcine et al.]{M. Mouhcine$^1$\thanks{Isaac Roberts Fellow}, 
S.P. Bamford$^2$\thanks{Current address: Institute of Cosmology and 
Gravitation, University of Portsmouth, Mercantile House, Hampshire Terrace,
Portsmouth, PO1 2EG, UK.}, A. Arag{\'o}n-Salamanca$^2$, O. Nakamura$^2$\\
$^{1}$Astrophysics Research Institute, Liverpool John Moores University,
Twelve Quays House, Egerton Wharf, Birkenhead, CH41 1LD, UK. \\
$^{2}$School of Physics and Astronomy, University of Nottingham,
       Nottingham, NG7 2RD, UK.
}

\date{Accepted ?.
      Received ?;
      in original form ?}

\pagerange{\pageref{firstpage}--\pageref{lastpage}} \pubyear{2004}

\begin{document}

\maketitle

\label{firstpage}

\begin{abstract}
  
We examine the evolutionary status of luminous, star-forming galaxies 
in intermediate-redshift clusters by considering their star formation 
rates and the chemical and ionsiation properties of their interstellar 
emitting gas.  Our sample consists of $17$ massive, star-forming, 
mostly disk galaxies with $M_{B} \la -20$, in clusters with redshifts 
in the range $0.31 \la z \la 0.59$, with a median of 
$\left\langle z \right\rangle = 0.42$. We compare these galaxies with 
the identically selected and analysed intermediate-redshift field 
sample of Mouhcine et al. (2006), and with local galaxies from the 
Nearby Field Galaxy Survey of Jansen et al. (2000).
  
From our optical spectra we measure the equivalent widths of {\oii}, 
{\hbeta} and {\oiiis} emission lines to determine diagnostic line 
ratios, oxygen abundances, and extinction-corrected star formation 
rates. The star-forming galaxies in intermediate-redshift clusters 
display emission line equivalent widths which are, on average, 
significantly smaller than measured for field galaxies at comparable 
redshifts. However, a contrasting fraction of our cluster galaxies 
have equivalent widths similar to the highest observed in the field. 
This tentatively suggests a bimodality in the star-formation rates 
per unit luminosity for galaxies in distant clusters.  We find no 
evidence for further bimodalities, or differences between our cluster 
and field samples, when examining additional diagnostics and the 
oxygen abundances of our galaxies. This maybe because no such 
differences exist, perhaps because the cluster galaxies which still 
display signs of star-formation have recently arrived from the field.  
In order to examine this topic with more certainty, and to further 
investigate the way in which any disparity varies as a function of 
cluster properties, larger spectroscopic samples are needed.

\end{abstract}

\begin{keywords}
galaxies: evolution -- galaxies: abundances --
galaxies: fundamental parameters -- galaxies: clusters: general
\end{keywords}

\section{Introduction}

One of the central problems in astronomy is that of galaxy formation and
evolution: when were the visible parts of galaxies assembled, when were 
the stars formed, and how did this depend on environment?

At low redshift the dependence of galaxy properties on environment is well
established.  Dense environments contain a larger fraction of red, passive
galaxies, while low density environments show more blue, star-forming 
galaxies (e.g., Lewis et al. 2002; G\'omez et al. 2003, Balogh et al. 2004).
Furthermore, star-forming cluster galaxies tend to have lower rates of
star formation than those in the field (Moss \& Whittle 1993, 2000).  These
star formation patterns are accompanied by a morphological trend, such that
denser regions have a larger fraction of early-types than the field (Morgan
1961, Dressler 1980).  However, the origin of these environmental dependences
is not yet clear, and continues to be debated.

As well as the local variation in galaxy properties with environment, there
are differences between the galaxy populations observed in comparable
environments at different cosmic epochs.  In intermediate-redshift clusters
there are a higher fractions of blue (Butcher \& Oemler 1978, 1984),
star-forming (Couch \& Sharples 1987, Dressler \& Gunn 1992, Poggianti 
et al. 1999), and spiral (Dressler et al. 1997) galaxies than in comparable 
local clusters.
The integrated star formation rate per cluster mass is thus larger for higher 
redshift clusters (Finn et al. 2005, Homeier et al. 2005). 
However, the field also evolves substantially with redshift (e.g., Cowie et
al. 1997), and intermediate-redshift cluster galaxies still always tend to 
have lower star formation rates (Balogh et al. 1998), star-forming fractions
(Postman et al. 2001, Poggianti et al. 2006) and spiral fractions
(Dressler et al. 1997) than field galaxies at the same epoch.

Recently the Tully-Fisher relation (Tully \& Fisher 1977) has been used to
investigate the differential evolution of cluster versus field galaxies at
intermediate redshift.  This method effectively utilizes rotation velocity, a
proxy for total mass, as a baseline against which other galaxy properties, in
this case luminosity, can be compared for different samples.  Ziegler et al.
(2003) have concluded that cluster galaxies at $z=0.3$--$0.5$ are distributed
in the Tully-Fisher diagram similarly to field galaxies at the same redshifts,
suggesting that the mass-to-light ratios of cluster galaxies are not affected
by physical processes specific to high density regions.  In contrast,
Milvang-Jensen et al. (2003) found mild evidence that cluster galaxies are
brighter than field ones at a fixed rotation velocity.  Using a larger sample,
Bamford et al.  (2005) also found that for luminous, star-forming field and
cluster disc galaxies at $0.3 \la z \la 0.8$, the cluster galaxies are
systematically brighter with respect to the Tully-Fisher relation of the 
field at the $3 \sigma$ level.  They speculate that this might be due to 
the increased presence of young stellar populations in galaxies which have
recently entered the cluster environment, due to an enhancement of star
formation caused by some interaction with that environment. However, Nakamura
et al. (2006) fail to confirm this cluster--field offset with a comparable
sample.  The question is therefore not yet settled, and awaits the results
from much larger studies, such as the ESO Distant Cluster Survey (EDisCS;
White et al. 2005) and the Deep Extragalactic Exploratory Probe 2 (DEEP2;
Davis et al. 2003).

The degree of chemical enrichment in distant star-forming galaxies provides
insight into how populations of distant galaxies map onto those in the local
universe. Analyses of the oxygen abundances of star-forming field galaxies at
intermediate redshifts seem to indicate that the luminosity--metallicity
relation evolves with redshift, with steeper slope (faster variation in
metallicity with luminosity) at earlier cosmic times (Kobulnicky et al. 2003;
Maier et al. 2004; Liang et al. 2004).  These studies imply that lower
luminosity field galaxies have experienced substantial chemical evolution
since $z \sim 1$, while the brightest galaxies have changed little.  Mouhcine
et al. (2006) found that the properties of the interstellar star-forming gas
for a sample of luminous, massive field galaxies at $0.2 \la z \la 0.8$ cover
a wide range, extending from those observed for local bright galaxies to those
of local dwarf galaxies. A subsample of these galaxies have already undergone
significant chemical enrichment, as indicated by their high oxygen abundances.
However, at a given galaxy luminosity many field galaxies have oxygen
abundances, ${\doh} \sim 8.6$, significantly lower than local galaxies with
similar luminosities.  These galaxies exhibit physical conditions, i.e.,
emission line equivalent width and ionization state, very similar to those of
local {\it faint}, metal-poor, star-forming galaxies.

In this paper, we analyze star formation rates and the properties of the
interstellar emitting gas for a sample of $17$ intermediate redshift cluster
galaxies.  Seven of these have securely measured rotation velocities and
emission scale lengths. The galaxies are drawn from the Bamford et al. (2005, 
2006) and Nakamura et al. (2006) samples, which were primarily constructed 
to compare the Tully-Fisher relations of cluster and field galaxies at 
intermediate redshifts.  We measure emission line equivalent widths,
diagnostic ratios, and oxygen abundances for cluster galaxies, with the aim of
investigating the environmental effects on the chemical content of galaxies at
intermediate redshift. Even with a small sample of cluster galaxies, we are
able to discern clear differences between intermediate redshift field and
cluster galaxy properties.

The paper is organized as follows. In Section~\ref{data} we briefly describe
the sample selection and emission line measurements.  In Section~\ref{prop},
we discuss the differences in ionization condition, luminosity--metallicity
relation, and star formation rates for cluster galaxies at intermediate
redshifts with respect to both intermediate redshift and local field samples.
Our results are summarised, and their implications briefly discussed, in
Section~\ref{disc}.

A concordance cosmological model with ${\rm H_0=70\,km\,s^{-1}}$,
$\Omega_{\Lambda}=0.7$, $\Omega_{m}=0.3$ has been adopted throughout the
paper.

\section{Observations and sample selection}
\label{data}

\begin{table}
 \caption{Coordinates and membership of cluster galaxies in our 
sample}
 \label{models_tab}
 \begin{tabular}{@{}lccccccc}
\hline
   ID & R.A. (J2000) & Dec. (J2000) && Cluster \\
 \hline

1  & 02:39:54.4 & -01:33:35  & A~370          \\
2  & 02:39:46.4 & -01:32:17  & A~370          \\
3  & 02:39:59.2 & -01:35:03  & A~370          \\
4  & 22:58:34.0 & -34:46:52  & AC~114         \\
5  & 22:58:40.6 & -34:50:12  & AC~114         \\
6  & 22:58:46.3 & -34:46:43  & AC~114         \\
7  & 22:58:49.3 & -34:47:01  & AC~114         \\
8  & 00:56:48.4 & -27:40:03  & CL~0054-27     \\
9  & 20:56:22.7 & -04:35:54  & MS~2053.7-0449 \\
10 & 00:18:31.1 &  16:22:05  & MS~0015.9+1609 \\
11 & 00:18:26.2 &  16:25:08  & MS~0015.9+1609 \\
12 & 00:18:30.9 &  16:25:41  & MS~0015.9+1609 \\
13 & 00:18:29.2 &  16:23:12  & MS~0015.9+1609 \\
14 & 16:23:42.3 &  26:30:55  & MS~1621.5+2640 \\
15 & 16:23:39.1 &  26:36:15  & MS~1621.5+2640 \\
16 & 16:23:38.0 &  26:35:40  & MS~1621.5+2640 \\
17 & 16:23:41.5 &  26:35:37  & MS~1621.5+2640 \\
 \hline
\end{tabular}
\end{table}

\setlength{\tabcolsep}{1.5pt}
\begin{table*}
 \centering
 \begin{minipage}{168mm}
  \caption{\label{tab1}
Properties of our sample of intermediate redshift cluster galaxies.
The columns give the ID, redshift, absolute rest-frame $B$-band magnitude, 
rest-frame equivalent widths of {\oii}, {\hbeta} and {\oiiis}, rotation 
velocity, emission scalelength, star formation rate determined from {\hbeta}, 
and the colour excess due to internal dust extinction, respectively.}
  \begin{tabular}{@{}lccccccccccc@{}}
  \hline
ID & $z$ & $M_{B}$ & EW({\oii}) & EW({\hbeta}) & EW({\oiiis}) & $V_{rot}$         & size  & {\doh} & SFR() & E(B-V)\\
   &     & mag   & (\AA)  & (\AA)  & (\AA)  & (${\rm km s^{-1}}$) & (kpc) & & (${\rm M_{\odot} yr^{-1}})$ & (mag) \\
 \hline
1  & 0.373 &  -22.03 $\pm$ 0.13 & 16.6 $\pm$ 0.5 & 12.8 $\pm$ 0.3 &  1.0 $\pm$ 0.2 & ...                 & ...                 & 9.02 $\pm$ 0.01 & 9.23 $\pm$ 0.23 & 0.12  $\pm$ 0.04 \\
2  & 0.378 &  -21.13 $\pm$ 0.17 & 30.3 $\pm$ 1.8 &  5.1 $\pm$ 0.9 &  4.9 $\pm$ 0.7 & 247. $\pm$ 11.      & 7.7 $\pm$ 0.3       & 8.41 $\pm$ 0.21 & 1.62 $\pm$ 0.28 & 0.000 $\pm$ 0.31 \\
3  & 0.384 &  -20.46 $\pm$ 0.1  & 26.8 $\pm$ 1.5 &  6.3 $\pm$ 0.6 &  7.2 $\pm$ 0.6 & ...                 & ...                 & 8.58 $\pm$ 0.09 & 1.08 $\pm$ 0.10 & 0.000 $\pm$ 0.18 \\
4  & 0.306 &  -20.72 $\pm$ 0.08 & 13.7 $\pm$ 0.9 &  2.1 $\pm$ 0.3 &  1.6 $\pm$ 0.4 & $81._{+23.}^{-16.}$ & $4.3_{+0.2}^{-0.3}$ &  ...            &         ...     & ...              \\       
5  & 0.313 &          ...       & 19.6 $\pm$ 1.8 &  7.4 $\pm$ 1.  &  5.6 $\pm$ 0.8 & ...                 & ...                 & 8.79 $\pm$ 0.07 &        ...      & ...              \\       
6  & 0.312 &  -21.09 $\pm$ 0.08 & 12.2 $\pm$ 0.6 &  3.5 $\pm$ 0.2 &  4.8 $\pm$ 0.2 & ...                 & ...                 & 8.64 $\pm$ 0.04 & 1.07 $\pm$ 0.05 & 0.14  $\pm$ 0.09 \\
7  & 0.313 &  -21.26 $\pm$ 0.08 &  6.7 $\pm$ 0.9 &  4.4 $\pm$ 0.6 &  4.3 $\pm$ 0.2 & $85._{+15.}^{-13.}$ & 3. $\pm$ 0.6        & 8.89 $\pm$ 0.04 & 1.56 $\pm$ 0.23 & 0.51  $\pm$ 0.22 \\
8  & 0.559 &  -21.06 $\pm$ 0.12 & 26.2 $\pm$ 1.8 & 22.7 $\pm$ 1.5 &  3.5 $\pm$ 1.6 & 194. $\pm$ 35.      & 2.5 $\pm$ 0.4       & 9.03 $\pm$ 0.01 & 6.74 $\pm$ 0.44 & 0.21  $\pm$ 0.11 \\
9  & 0.588 &  -20.38 $\pm$ 0.1  & 63.8 $\pm$ 3.8 & 25.6 $\pm$ 2.7 & 16.2 $\pm$ 1.7 & ...                 & ...                 & 8.83 $\pm$ 0.05 & 4.07 $\pm$ 0.43 & 0.000 $\pm$ 0.16 \\
10 & 0.551 &  -21.34 $\pm$ 0.1  & 11.9 $\pm$ 1.3 &  3.2 $\pm$ 1.0 &  4.1 $\pm$ 0.9 & ...                 & ...                 & 8.63 $\pm$ 0.26 & 1.24 $\pm$ 0.39 & 0.21  $\pm$ 0.53 \\
11 & 0.554 &  -20.89 $\pm$ 0.09 &  7.6 $\pm$ 2.8 &  5.2 $\pm$ 1.3 &  3.2 $\pm$ 1.6 & ...                 & ...                 & 8.94 $\pm$ 0.05 & 1.32 $\pm$ 0.33 & 0.69  $\pm$ 0.47 \\
12 & 0.549 &  -22.28 $\pm$ 0.08 & 23.9 $\pm$ 0.9 &  4.1 $\pm$ 0.8 &  2.2 $\pm$ 0.5 & ...                 & ...                 & 8.43 $\pm$ 0.24 & 3.67 $\pm$ 0.73 & 0.000 $\pm$ 0.35 \\
13 & 0.550 &  -20.44 $\pm$ 0.11 & 19.6 $\pm$ 1.9 &  6.9 $\pm$ 1.2 &  5.3 $\pm$ 2.5 &$137._{+49.}^{-37.}$ & $4.6_{+0.5}^{-0.4}$ & 8.77 $\pm$ 0.11 & 1.16 $\pm$ 0.20 & 0.10  $\pm$ 0.30 \\
14 & 0.424 &  -21.15 $\pm$ 0.07 & 10.5 $\pm$ 1.0 &  4.9 $\pm$ 0.8 &  5.4 $\pm$ 0.4 & $213._{+7.}^{-6.}$  & $2.7_{+0.1}^{-0.2}$ & 8.81 $\pm$ 0.07 & 1.56 $\pm$ 0.26 & 0.49  $\pm$ 0.25 \\
15 & 0.422 &  -20.93 $\pm$ 0.09 & 63.2 $\pm$ 1.6 & 21.7 $\pm$ 1.  & 19.6 $\pm$ 0.7 & ...                 & ...                 & 8.75 $\pm$ 0.03 & 5.71 $\pm$ 0.25 & 0.000 $\pm$ 0.07 \\
16 & 0.421 &  -20.75 $\pm$ 0.08 & 24.5 $\pm$ 1.6 &  7.9 $\pm$ 0.9 &  6.9 $\pm$ 0.7 &$216._{+19.}^{-16.}$ & $3.2_{+0.3}^{-0.3}$ & 8.73 $\pm$ 0.08 & 1.76 $\pm$ 0.21 & 0.09  $\pm$ 0.19 \\
17 & 0.429 &  -19.97 $\pm$ 0.08 & 14.1 $\pm$ 2.7 &  6.2 $\pm$ 1.5 &  3.3 $\pm$ 1.3 & ...                      & ...                   & 8.86 $\pm$ 0.09 & 0.67 $\pm$ 0.16 & 0.09  $\pm$ 0.38 \\

\hline
\end{tabular}
\end{minipage}
\end{table*}

The observed galaxies are located in nine fields, centered on the 
clusters MS~0440.5+0204 (${\rm \zcl=0.20, \sigmacl=838 km s^{-1}}$), 
A~2390 (${\rm \zcl=0.23, \sigmacl=1294 km s^{-1}}$), 
AC~114 (${\rm \zcl=0.32, \sigmacl=1388 km s^{-1}}$), 
A~370 (${\rm \zcl=0.37, \sigmacl=859 km s^{-1}}$), 
MS~1621.5+2640 (${\rm \zcl=0.43, \sigmacl=735 km s^{-1}}$), 
MS~0015.9+1609 (${\rm \zcl=0.55, \sigmacl=984 km s^{-1}}$), 
CL~0054-27 (${\rm \zcl=0.56, \sigmacl=742 km s^{-1}}$), 
MS~2053.7-0449 (${\rm \zcl=0.58, \sigmacl=817 km s^{-1}}$), and 
MS~1054.4-0321 (${\rm \zcl=0.83, \sigmacl=1178 km s^{-1}}$), 
where $\zcl$ and $\sigmacl$ cluster redshifts and velocity-dispersions, 
obtained from the literature (Stocke et al. 1991; Girardi \& Mezzetti 
2001; Hoekstra et al. 2002). Spectroscopy and imaging were obtained 
using FORS2\footnote{\label{fn:fors}http://www.eso.org/instruments/fors} 
on the VLT (Seifert et al. 2000), for MS~0440.5+0204, AC~114, A~370, 
CL~0054-27, MS~2053.7-0449, MS~1054.4-0321, and using 
FOCAS\footnote{\label{fn:focas}http://www.naoj.org/Observing/Instruments/FOCAS}
on the Subaru telescope (Kashikawa et al. 2002) for A~2390, 
MS~1621.5+2640, MS~0015.9+1609, MS~2053.7-0449 again. The cluster 
galaxies considered in this paper are defined as those galaxies with 
redshifts, $z$, such that $\zcl - 3\sigmacl \le z \le \zcl + 3 \sigmacl$.
Galaxies with redshifts outside of these limits were assigned to the 
field sample considered in Mouhcine et al. (2006).

As the original samples were designed to study the Tully-Fisher relation, 
the galaxies observed within each field were selected by assigning 
priorities based upon the likelihood of being able to measure a rotation 
curve. Galaxies with disky morphologies, favourable inclination for 
rotation velocity measurement, known emission line spectrum, and 
available Hubble Space Telescope data were assigned higher priorities. 
The sample is thus biased toward luminous, star-forming disc galaxies, 
and therefore we are not probing the average galaxy population in 
clusters.

The reduction of the spectroscopic and photometric data are detailed 
in Bamford et al. (2005) and Nakamura et al. (2006), and summarised 
in Mouhcine et al. (2006). The products utilized in this paper are 
wavelength-calibrated, sky-subtracted, non-flux-calibrated 1d-spectra
spanning the observed range $\sim 5000$--$8500$ {\AA} at $4.2$--$6.3$ 
{\AA} resolution (FWHM), from spatial regions covering the whole major 
axis of each galaxy, and absolute rest-frame $B$-band total magnitudes.

To measure emission line equivalent widths, single Gaussian fits were
attempted for all visible emission lines. Balmer emission lines were 
corrected for the underlying stellar absorption by considering 
simultaneous fits of the emission and absorption lines. For a fraction 
of the galaxy spectra, particularly with low \snr{}, these two-component 
fits were found to be unreliable. To correct the affected galaxies, we 
thus apply a uniform stellar absorption correction estimated from the 
galaxies with both reliable two- and one-component fits (see Mouhcine 
et al. 2006 for more details).

Our initial sample of cluster galaxies with identifiable emission lines
contains $72$ galaxies, and spans a redshift range of $0.2$ to $0.84$ 
with a median of $0.55$. We searched within this spectroscopic sample 
for galaxies with emission lines suitable for chemical analysis. 
Only galaxies for which it was possible to measure {\oii}, {\hbeta}, 
and {\oiiis} emission lines were retained, i.e., the lines needed to 
determine the ionizing source and to measure the oxygen abundance. 
After applying these selection criteria the sample size drops to $22$. 
For reliable oxygen abundance determinations, only galaxies for which 
{\hbeta} emission line is well detected are retained. This was judged 
by requiring the {\snr}, estimated from the median pixel value in 
regions $29$--$58$\AA{} away from the line on both sides, divided by 
the median value of the error image in the same region, to be larger 
than 8. An additional $5$ objects were excluded from the sample due to 
the weak detection of {\hbeta}. Of the the original data set with $72$ 
objects, the final sample contains $17$ emission line galaxies in the 
redshift range $0.31 \la z \la 0.59$, with a median of 
$\left< z \right> = 0.42$.  These galaxies span almost 2.5 magnitude 
in $B$-band luminosity ($M_B = -19.9$ to $-22.3$), with a median of 
$-21$. Seven of the cluster galaxies in our sample have securely 
measured rotation velocities, in the range $\vrot = 82$--$248$ 
kms$^{-1}$ with a median of $194$ kms$^{-1}$, and emission scale 
length in the range $\rdspec = 2.5$--$7.7$ kpc, with a median of 
$3.2$ kpc (see Bamford et al. 2005 for more details on the 
measurements of rotation velocities and emission scale lengths).
\footnote{These numbers reduce to $16$ galaxies with oxygen abundances
($6$ of which have measured $\vrot$ \& $\rdspec$), after we reject one
galaxy due to difficulties in measuring its metallicity, and $15$
galaxies with star formation rates and internal dust reddenings ($6$
with $\vrot$ \& $\rdspec$) after we reject another galaxy with no
magnitude measurement; see section \ref{comp}}

\begin{figure}
\centering
\includegraphics[clip=,angle=0,width=0.45\textwidth]{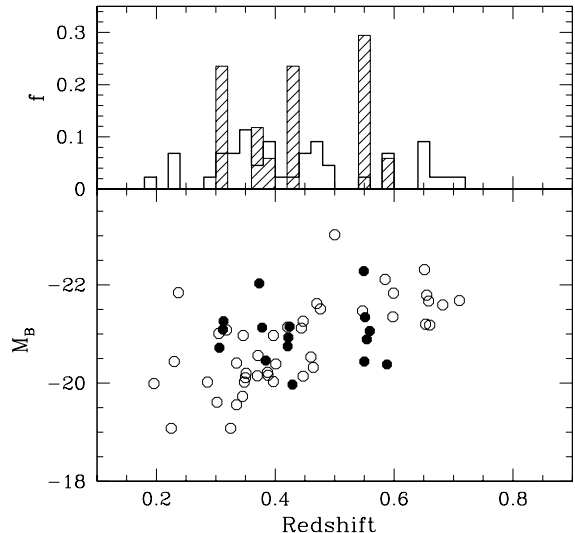}
\caption{Lower panel: The relationship between B-band absolute magnitude 
and redshift for the sample of cluster galaxies, shown as field circles. 
The relationship for the comparison sample of field galaxies is shown as 
open circles. Upper panel: The redshift distribution for field and cluster 
galaxy samples shown are open and hatched histograms respectively.}
\label{fig:mb_z_distz}
\end{figure}

The lower panel of Fig.~\ref{fig:mb_z_distz} shows the relationship 
between rest-frame B-band absolute magnitude and redshift for the 
final cluster galaxy sample, shown as filled circles, and a comparison 
sample of intermediate redshift field galaxies from Mouhcine et al.  
(2006). The figure shows that the redshift and the galaxy luminosity 
are correlated for the comparison sample of field galaxies. No clear 
correlation is seen for our cluster galaxies, however the luminosities 
covered at a given redshift are consistent for the two samples. 

The upper panel of the figure shows the redshift distribution for 
galaxies in our intermediate redshift cluster (hatched, thin line) and 
field (thick line) samples. Clearly, the bulk of field galaxies in the 
comparison sample are distributed over a redshift interval similar to
our cluster galaxies. The redshift distribution of the field galaxies
does, however, extend beyond that of our cluster galaxies.

For the high-redshift end of the distribution, we lose the {\oiiis} 
emission line from the spectral range above $z \sim 0.7$, so that the 
final sample does not contain any galaxy from the most distant cluster
(MS~1054.4-0321). Note that there is not a sharp redshift cutoff at 
which we lose the required lines, as the exact wavelength range of the 
spectra depend upon the position of the slit in the mask (see Bamford 
et al. 2005 and Nakamura et al. 2006 for more details).

A similar situation occurs at the low-redshift end of the distribution. 
We have a few field galaxies at $z \sim 0.2$, but no cluster galaxies 
due to the slightly different wavelength ranges of the VLT and Subaru 
data. For the VLT data the cutoff is setting in by $z \approx 0.2$, the 
redshift of MS0440. For Subaru data the cutoff occurs above $z=0.23$, 
which is the reason why the final sample does not contain any galaxies 
from cluster A2390. We are also missing field galaxies at redshifts 
around the two clusters at $\zcl=0.55$ and $\zcl=0.56$ due to the 
presence of telluric atmospheric absorption at $\sim 7580$--$7700$ \AA. 
The spectra were not corrected for this feature, and so the lines which
fell into this region were not used due to the difficulty in measuring
their equivalent widths. This occurs for {\oiiis} at $z=0.514-0.538$
and for {\hbeta} at $z=0.559-0.584$. The cluster galaxy at $z \sim
0.559$ is just outside the affected range.
It is worth mentioning that the results presented in this paper do not
change if the field galaxies beyond the limits where cluster galaxies 
are detected are excluded from the intermediate redshift comparison 
sample.

An accurate estimate of the interstellar star-forming gas properties 
requires that the observed emission lines arise in {H{\sc ii}} regions, 
powered by photoionization from massive stars.  None of the observed 
spectra contain all the required emission lines to determine the 
ionizing source using the classical techniques. Therefore, we use the 
equivalent width ratios of {\oiii}/{\hbeta} and {\oii}/{\hbeta} as 
parametrized by Lamareille et al. (2004) to check the nature of the 
ionizing source.  All the objects in our sample fall within the zone 
where starburst galaxies are located, indicating that for all of them 
the source ionizing the interstellar gas is an episode of star 
formation.

\section{Properties of cluster galaxies}
\label{prop}

\begin{table*}
 \centering
 \begin{minipage}{157mm}
  \caption{\label{tab:param}
Statistical comparison of our cluster and field samples.
For each parameter ($x$), the columns give the number of cluster and field
objects ($n_{cl}$ and $n_{f}$ respectively), the robust mean of that
parameter for cluster and field objects ($\overline{x}_{cl}$ and
$\overline{x}_{f}$ respectively), the cluster$-$field difference in the
means ($\Delta(\overline{x}_{cl} - \overline{x}_{f})$), an estimate of the
size of cluster$-$field difference that would be required in order to
discriminate between the two at the $3\sigma$-level ($|\Delta_{dis}| =
3\times\sqrt(\sigma_{\overline{x}_{cl}}^2+\sigma_{\overline{x}_{f}}^2)$), 
the probability that the cluster and field populations have the same mean
($P(\mu_{cl}\!=\!\mu_{f})$), and the KS-test probability that the cluster
and field population distributions are the same ($P(\rmn{KS})$).  }
  \begin{tabular}{rccr@{$\,\pm\,$}lr@{$\,\pm\,$}lcccc}
  \hline
    $x$ & $n_{cl}$ & $n_{f}$ & \multicolumn{2}{c}{$\quad\overline{x}_{cl}$} &
    \multicolumn{2}{c}{$\quad\overline{x}_{f}$} & $\Delta(\overline{x}_{cl} - \overline{x}_{f})$ &
    $|\Delta_{dis}|$ & $P(\mu_{cl}\!=\!\mu_{f})$ \% & $P(\rmn{KS})$ \%\\
 \hline
  EW({\oii}) (\AA)            & 17 & 44 &                  $   17.55 $  & $   2.68 $ &
$   33.86 $ & $   2.82 $ & $-16.31 $ & $   11.68 $ & $  0.013^{\rmn{a}} $ & $  0.991 $\\
  EW({\hbeta}) (\AA)          & 17 & 44 &                  $    5.35 $  & $   0.65 $ &
$   10.24 $ & $   0.82 $ & $ -4.90 $ & $    3.14 $ & $  0.002^{\rmn{a}} $ & $  1.826 $\\
           \oo                & 17 & 44 &                  $    0.35 $  & $   0.05 $ &
$    0.33 $ & $   0.03 $ & $  0.02 $ & $    0.17 $ & $  75.63 $ & $  96.53 $\\
           \rp                & 17 & 44 &                  $    4.13 $  & $   0.48 $ &
$    4.87 $ & $   0.30 $ & $ -0.75 $ & $    1.71 $ & $  19.19 $ & $  44.96 $\\
       $M_{B}$ (mag)          & 16 & 43 & \makebox[0pt][r]{$  -20.90 $} & $   0.15 $ &
\makebox[0pt][r]{$  -20.79 $} & $   0.14 $ & $ -0.11 $ & $    0.61 $ & $  57.34 $ & $  19.87 $\\
SFR ($\rm M_{\odot}\,yr^{-1}$) & 15 & 39 &                  $    1.34 $  & $   0.23 $ &
$    2.37 $ & $   0.38 $ & $ -1.03 $ & $    1.32 $ & $  2.242^{\rmn{b}} $ & $  46.45 $\\
          \doh                & 16 & 40 &                  $    8.77 $  & $   0.05 $ &
$    8.71 $ & $   0.03 $ & $  0.06 $ & $    0.17 $ & $  25.09 $ & $  50.55 $\\
      E(B$-$V) (mag)          & 15 & 39 &                  $    0.10 $  & $   0.05 $ &
$    0.08 $ & $   0.01 $ & $  0.03 $ & $    0.16 $ & $  61.53 $ & $  74.35 $\\
\hline
\end{tabular}
\\
$^{\rmn{a}}$ highly significant\quad $^{\rmn{b}}$ significant, but see
discussion in text for caveats.
\end{minipage}
\end{table*}

\subsection{Comparison samples \& oxygen abundances}
\label{comp}

We take a step toward quantifying environmental variation in galaxy 
properties at intermediate redshifts by comparing our cluster galaxy 
properties with those of field galaxies from Mouhcine et al.  (2006). 
The redshift distributions of both field and cluster galaxy samples are 
comparable, the field galaxies span a redshift range $0.2 \la z \la 0.8$ 
with a median of $\left< z \right> = 0.45$.  The comparison sample of 
intermediate redshift field galaxies was selected and analysed in an 
identical manner to the sample studied in this paper.

To investigate differences between the properties of $z \sim 0.5$ cluster
galaxies and the general present day galaxy population, we require a local
comparison sample.  For this we use the sample of Jansen et al. (2000), 
who observed the Nearby Field Galaxy Survey (NFGS) of about 200 galaxies. 
The NFGS was selected from the first CfA redshift catalogue (Huchra et al. 
1983) to approximate the local galaxy luminosity function.  To select 
star-forming galaxies in the original NFGS sample, we use the classical 
diagnostic ratios of two pairs of relatively strong emission lines 
(Baldwin et al. 1981; Veilleux \& Osterbrock 1987). The colour excess 
from obscuration by dust for NFGS sample galaxies was estimated from the 
observed ratio of {\halpha} and {\hbeta} line fluxes. We adopt the Milky 
Way interstellar extinction law of Cardelli, Clayton, \& Mathis (1989), 
with $R_{V}=3.1$, and assume an intrinsic Balmer decrement of $2.85$, 
corresponding to the case B recombination with a temperature of 
${\rm T=10^4 K}$ and a density of ${\rm n_{e}\sim\,10^2-10^4\,cm^{-2}}$ 
(Osterbrock 1989).

Gas phase oxygen abundances are estimated using the so-called strong 
emission line method, based on measurements of {\oii}, {\oiii}, and 
{\hbeta} (McGaugh 1991). This is done through the parameter ${\rp} = 
({\oiii}+{\oii})/{\hbeta}$ introduced initially by Pagel et al. (1979). 
The {\rp} parameter is both abundance- and ionization-sensitive (e.g., 
Kewley \& Dopita 2002). The correction of this dependence of {\rp} 
parameter on ionization is usually done by using the ionization-sensitive 
diagnostic ratio {\oo}\,=\,{\oiii}/{\oii} (e.g., McGaugh 1991; 
Kewley \& Dopita 2002).  Note that the oxygen abundance estimated using 
different calibrations available in the literature might differ by factors 
up to $\sim 4$ (Ellison \& Kewley 2005), however, as our main interest is 
to study the relative change in oxygen abundances between cluster and 
field galaxies, the exact choice of the {\doh} vs. {\rp} calibration is 
not a critical issue. Here we determine oxygen abundance using the 
calibration of McGaugh (1991), as taken from in Kobulnicky, Kennicutt 
\& Pizagno (1999).

The use of {\rp} parameter to estimate oxygen abundance is complicated 
by the degenerate dependence of this parameter on metallicity: at the 
same value of {\rp}, different ionization parameters correspond to 
different oxygen abundances (McCall et al. 1985). The observed emission 
lines for our sample galaxy spectra are not enough to break the 
degeneracy. However, when the needed emission lines to do so are observed 
for intermediate redshift galaxies with similar luminosities, they are 
found to lie on the metal-rich branch of the {\doh} versus {\rp} 
calibration (Kobulnicky et al.  2003; Maier et al. 2005). To estimate 
the oxygen abundance, we make the assumption that the galaxies in our 
sample lie on the upper metallicity branch of the {\doh} versus {\rp}
calibration.  {\rp} and {\oo} have been estimated using emission line
equivalent widths. Kobulnicky \& Phillips (2003) have shown that 
estimates of both ratios using equivalent widths give results similar 
to using emission line fluxes.

For one galaxy in the sample, the oxygen abundance estimated using 
the low-branch of metallicity calibration, given its {\rp} and {\oo}
parameters, was larger than that derived using the upper-branch. 
This occurs when the measured {\rp} parameter reaches a higher value 
than the maximum allowed by the photoionization model, i.e., for a 
given {\oo}. The estimate of oxygen abundance in this case is highly 
uncertain. No estimate of the oxygen abundance was made for this 
galaxy.

None of the objects in our sample have both {\halpha} and {\hbeta} 
emission lines present in their covered spectral range, so the dust 
obscuration cannot be derived using the Balmer decrement. However, it 
is possible to estimate the amount of internal extinction by comparing 
the energy balance between the luminosities of two different star 
formation indicators, {\oii} and {\hbeta} in this case after applying 
the appropriate corrections, that are free from systematic effects 
other than dust reddening, i.e., that do not depend on the metallicity 
or excitation state of the emitting gas. The star formation rate in 
terms of the {\oii} emission line includes a correction factor to 
account for the effect of metallicity on the variation of the 
{\oii}/{\halpha} flux ratio, as calibrated by Kewley, Geller \& Jansen 
(2004; see Mouhcine et al. 2005).

Table~\ref{tab1} lists the redshifts, absolute $B$-band magnitudes, 
{\oii}, {\hbeta}, and {\oiiis} emission line rest-frame equivalent 
widths, rotation velocities, emission scale lengths, oxygen abundances, 
star formation rates, and colour excesses for the objects in our 
final sample of $17$ cluster galaxies.

\subsection{Results}

\begin{figure}
\centering
\includegraphics[clip=,angle=270,width=0.45\textwidth]{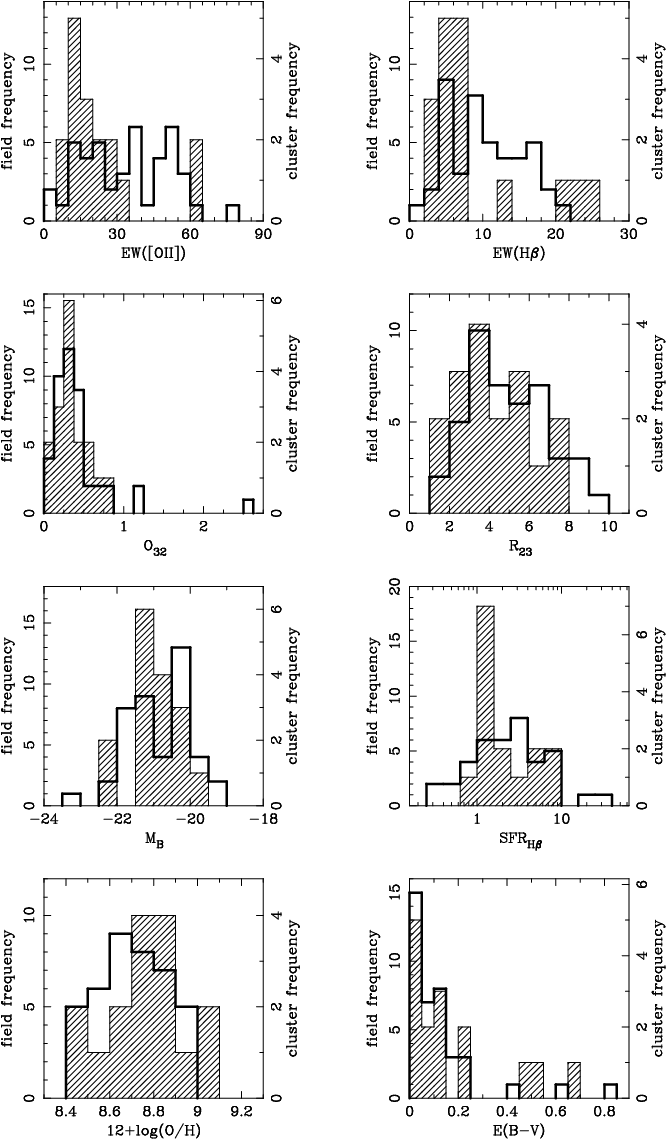}
\caption{The distributions of {\oii} and {\hbeta} equivalent width, 
the {\oo} and {\rp} parameters, $M_B$, star formation rate derived 
from {\hbeta} (shown as $\log_{10}{(\rmn{SFR}_{\rmn{H}\beta})}$ to aid 
comparison), oxygen abundance, and colour excess for galaxies in our 
intermediate redshift cluster (hatched, thin line) and field (thick 
line) samples. The cluster and field histograms are plotted on different 
scales such that they both enclose the same area.}
\label{fig:distrib}
\end{figure}

The distributions of {\oii} and {\hbeta} equivalent width, {\oo} and 
{\rp} parameters, $M_B$, star formation rate derived from {\hbeta}, 
oxygen abundance, and colour excess are shown for our intermediate 
redshift cluster and field samples in Fig.~\ref{fig:distrib}.  
In addition to the visual comparison between our cluster and field samples
afforded by these histograms, we have quantitatively compared the parameters.
This has been done primarily by estimating the difference between the
parameters' parent distribution means and evaluating the significance of 
this difference.  The results are given in table \ref{tab:param}. 
The figures presented in this table are from a comparison utilising robust 
biweight estimators (e.g., Beers, Flynn \& Gebhardt 1990). The probability 
that the parent distributions have the same mean is evaluated by a robust 
t-test, which does not assume that the distributions have equal variances 
(Welch 1937).  Similar results are found using canonical statistics, both 
unweighted and with weights corresponding to the measurement errors with
the inclusion of an intrinsic dispersion term.  An exception is
$\rmn{SFR}_{\rmn{H}\beta}$, for which the less robust statistics find a 
less significant difference between the cluster and field samples. 
Table \ref{tab:param} also lists the probability that the parent cluster 
and field distributions are the same as given by a Kolmogorov-Smirnov 
(KS) test.

The most striking differences are displayed by the equivalent widths of 
{\oii} and {\hbeta}, for which the cluster and field galaxy means differ 
at a $\ga\! 99.99$\% significance level.  The KS-test also 
finds differences at a $>\!98$\% significance level. The equivalent 
widths of {\oii} and {\hbeta} are on average significantly lower for 
cluster galaxies than for field galaxies in our samples, both by a factor 
of $\unisim 0.52\pm0.09$. If there is no variation in broad-band luminosities 
($M_B$) between these galaxy samples, these equivalent width ratios imply 
similar emission line luminosity ratios. Star formation rate is proportional 
to the {\hbeta} luminosity and, upto a metallicity and ionisation dependence, 
to the {\oii} luminosity. We therefore expect the star formation rates of 
our cluster galaxies to be on average lower than those in our field sample, 
by a similar factor.  Indeed, we find this to be true, by a factor of 
$0.6\pm0.1$.  However, this result is less significant than that based 
solely upon the equivalent widths.  The robust statistics presented in
table \ref{tab:param} find the difference to be $\unisim 98$\%, but canonical
statistics and the KS-test find no significant difference.  Also, if the
comparison is performed in the log-regime, which may be more appropriate, no
significant difference is found.  This reduction in significance is probably
due to broad-band luminosity variations. The spread in luminosity 
($\unisim 2$ mag, therefore a factor of $\unisim 6$ in luminosity) is 
comparable to the spread in equivalent width, and appears to blur out the 
differences between the cluster and field samples, reducing the significance 
of the difference, but not particularly affecting its magnitude.

\begin{figure*}
\centering
\includegraphics[clip=,width=0.45\textwidth]{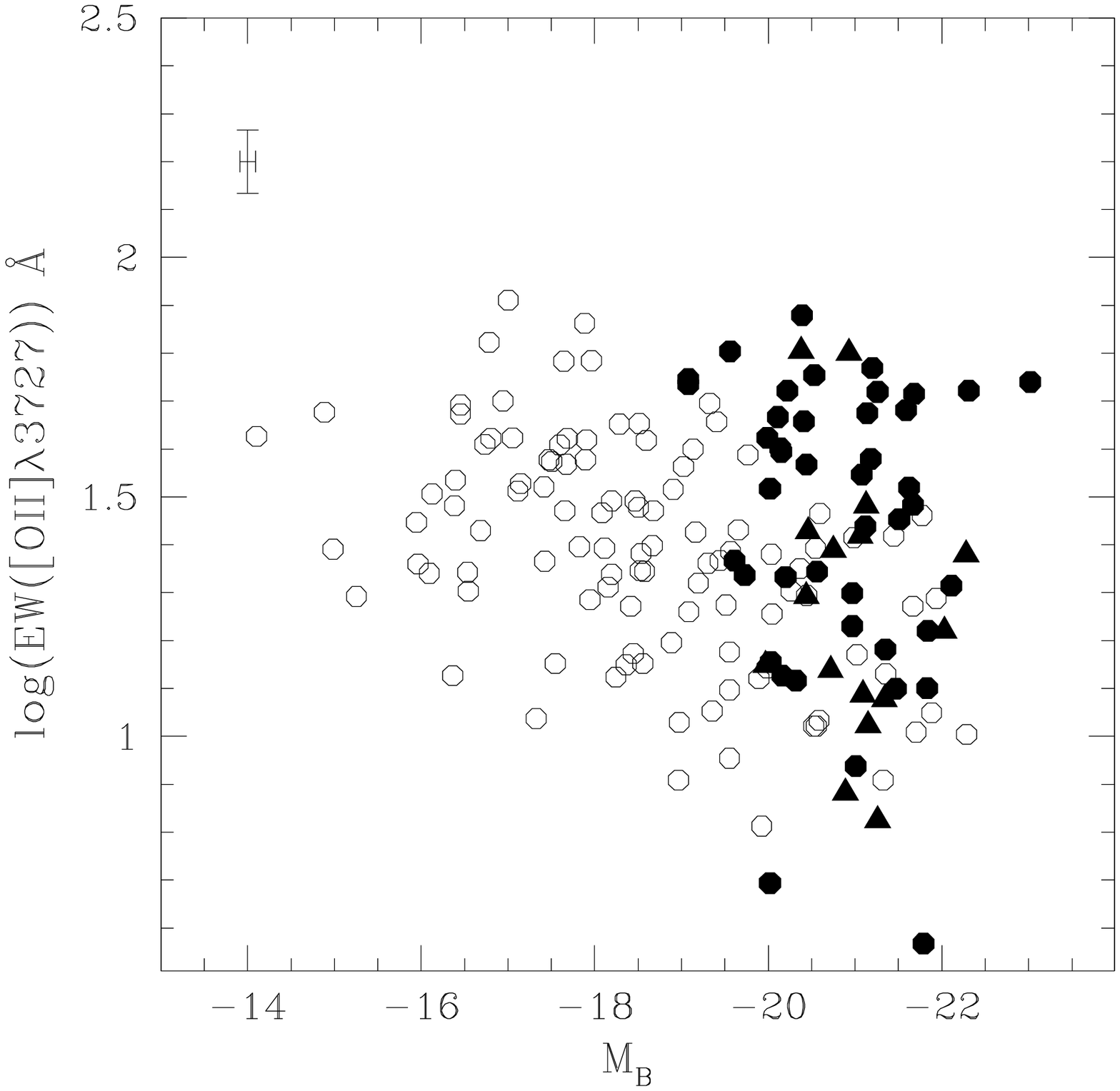}
\includegraphics[clip=,width=0.45\textwidth]{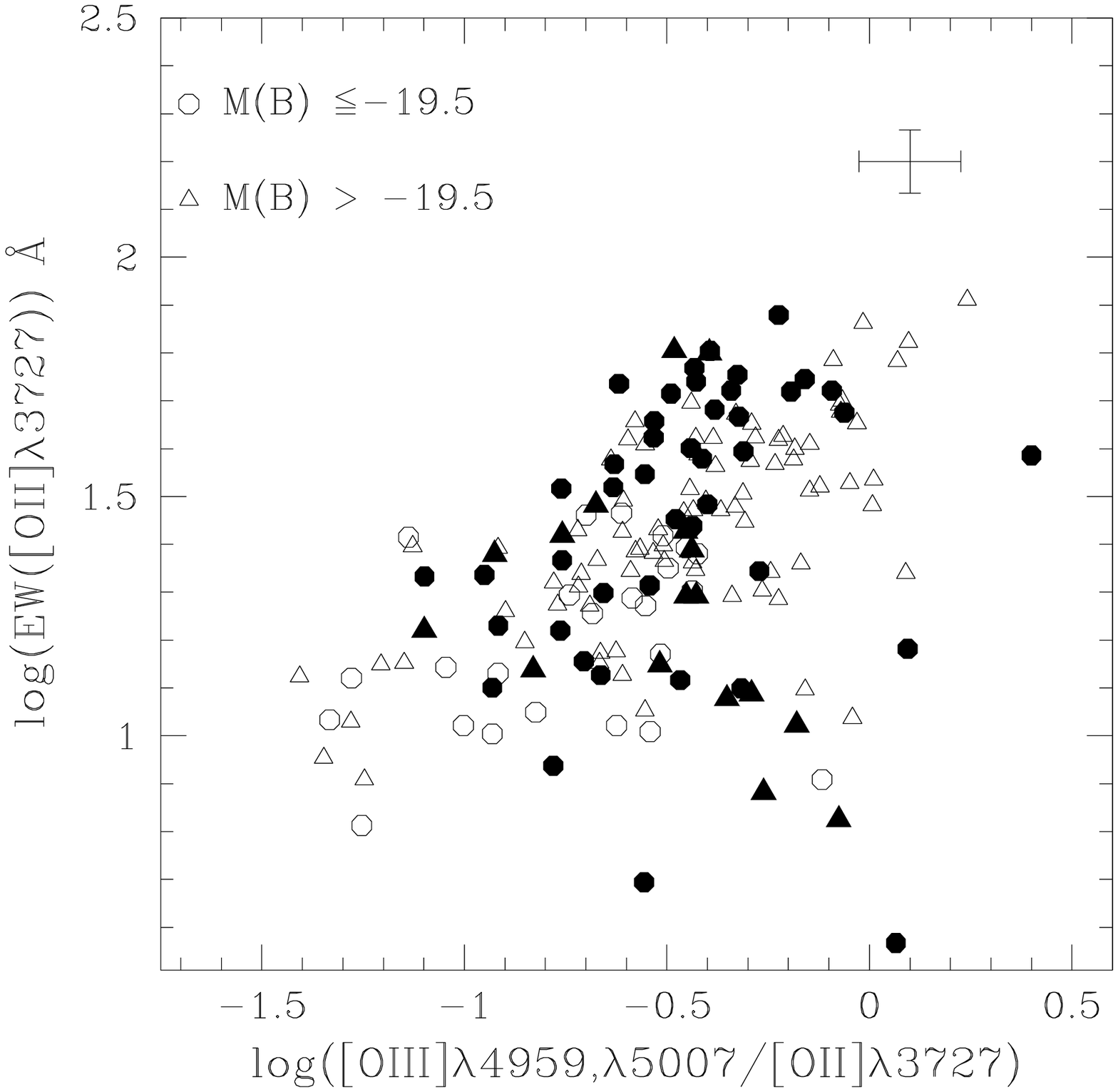}
\caption{{\it Left}: The relationship between rest-frame {\oii} emission 
line equivalent width and absolute $B$-band magnitude for our samples of 
intermediate redshift bright, star-forming galaxies (field galaxies as 
filled circles, and cluster members as filled triangles), and the NFGS 
local star-forming galaxies (open circles).  {\it Right}: Rest-frame 
{\oii} emission line equivalent width as a function of the
excitation-sensitive diagnostic ratio {\oo}.  Field star-forming galaxies 
at intermediate redshifts are shown as filled circles, and cluster members 
as filled traingles, open triangles show faint ($M_{B} > -19.5$) NFGS 
galaxies, and open circles show bright ($M_{B} > -19.5$) NFGS galaxies.  }
\label{ewo2_mb_o32}
\end{figure*}

\begin{figure}
\centering
\includegraphics[clip=,angle=270,width=0.40\textwidth]{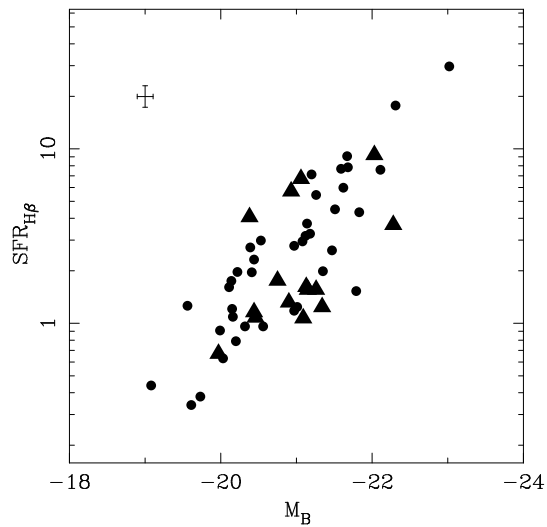}
\caption{Magnitude versus star formation rate, derived from {\hbeta}, 
for our samples of field (circles) and cluster (triangles) galaxies. 
Representative errors are shown by the error bars in the top left 
corner.}
\label{fig:mag_sfr}
\end{figure}

However, these statistics do not tell the whole story.  While most cluster
galaxies have equivalent widths on the low side of the field distrib, there
are several objects ($10$--$20$\%) which have equivalent widths as high as 
any of the field galaxies. This bimodality is visible in Fig.~\ref{ewo2_mb_o32}
in the plot of {\oii} equivalent width versus $B$-band magnitude. It also 
remains in the star formation rate distribution, with the majority of our 
cluster galaxies having star formation rates at the low end of the field 
distribution, but with several objects ($\unisim 20$\%) having SFRs higher 
than any of the field galaxies with the same broad-band luminosity. This is 
shown clearly by a plot of star formation rate versus magnitude, as given 
in Fig.~\ref{fig:mag_sfr}.  These results tentatively imply that distant 
cluster galaxies have a star formation rate per unit luminosity that is 
lower than the average for coeval field galaxies, with the exception of 
a subsample of $\unisim 20$\% cluster galaxies, which have star formation 
rates per unit luminosity that are higher than those usually seen in the 
field.  We do not find evidence for bimodalities in the distribution
 histograms for any of the other parameters measured for our distant field 
and cluster galaxies.

Locally, the strength of emission lines is known to correlate with other 
galaxy properties, e.g., luminosity, metallicity, and ionization conditions 
(e.g., McCall et al. 1985; Stasi\'nska 1990; Kewley \& Dopita 2002; Mouhcine 
et al. 2005). On average faint/metal-poor galaxies tend to be highly ionized, 
while luminous/metal-rich galaxies are characterized by low-ionization 
parameters. The left panel of Fig.~\ref{ewo2_mb_o32} shows the relationship 
between galaxy absolute $B$-band magnitude and {\oii} rest-frame emission 
line equivalent width. Our cluster galaxies are shown as filled triangles, 
the sample of intermediate
redshift field galaxies from Mouhcine et al. (2006) is shown by filled
circles, and the local star-forming galaxies in the NFGS sample are shown as
open circles.  The NFGS galaxies in this figure display the well-established
correlation between galaxy luminosity and emission line equivalent width
(e.g., Salzer et al. 1989; Kong et al. 2002; Jansen et al. 2000).  
As discussed in Mouhcine et al. (2006), the distant field galaxies in our
sample cover a similar range of {\oii} rest-frame emission line equivalent 
width to that observed locally, but over a much narrower luminosity range, 
that is, $\sim\!2$ mag in comparison to the $\sim\!7$ mag covered by the 
NFGS galaxies. Strikingly, intermediate redshift cluster galaxies seem 
to be mostly located in similar regions of the diagnostic diagrams as local 
luminous galaxies, while {\oii} equivalent widths for galaxies in the field 
extend to values observed locally only at much lower luminosities.  
As mentioned earlier, an exception to this trend is provided by two cluster 
galaxies with {\oii} equivalent widths higher than nearly all of our distant 
field galaxies.

The right panel of Fig.~\ref{ewo2_mb_o32} shows the variation of {\oii}
emission line rest-frame equivalent width as a function of the
ionization-sensitive diagnostic ratio {\oo}\,=\,{\oiii}/{\oii}. Intermediate 
redshift cluster and field galaxies are shown as in the left panel of the 
figure. To illustrate the effect of galaxy luminosity on {\oo}, we split 
the local sample of star-forming galaxies into faint ($M_{B} > -19.5$) and 
bright ($M_{B} \le -19.5$), samples. As discussed in Mouhcine et al. (2006), 
the bright, star-forming field galaxies at intermediate redshifts tend to 
be located in the same region as \emph{faint} local star-forming galaxies 
in the plot of {\oii} equivalent width versus {\oo}, and show higher
ionization-sensitive diagnostic ratios ({\oo}) than are seen locally in
galaxies with comparable luminosities. However, our distant, star-forming
cluster galaxies preferentially inhabit the same region in the {\oii} 
equivalent width versus {\oo} diagram as local field galaxies with similar, 
bright, luminosities. Exceptions to this are the same two high {\oii} 
equivalent width galaxies discussed previously, and the population of 
high-{\oo} cluster galaxies. While the {\oo} distributions of our distant 
field and cluster galaxies are similar (see Fig.~\ref{fig:distrib}), the 
{\oii} equivalent widths display a strong difference. The galaxies most 
responsible for this difference appear to be those with {\oo} $ \ga 0.4$.  
This combination of high {\oo} and low {\oii} equivalent widths is rather 
more unusual in both our distant and local field samples.
  
The broad similarity between the observed rest-frame emission line equivalent
widths and diagnostic ratios for bright, star-forming galaxies in the local
field and those in intermediate-redshift clusters, suggests that the two
samples span the same range of {H{\sc ii}} region physical parameters, in
terms of ionizing flux, ionization parameter, and metallicity.  This also
indicates that the {\rp} vs. {\doh} local calibration should be valid for
converting line ratios measured for distant cluster galaxies in our sample
into oxygen abundances, without introducing any systematic biases.

\begin{figure}
\centering
\includegraphics[clip=,width=0.45\textwidth]{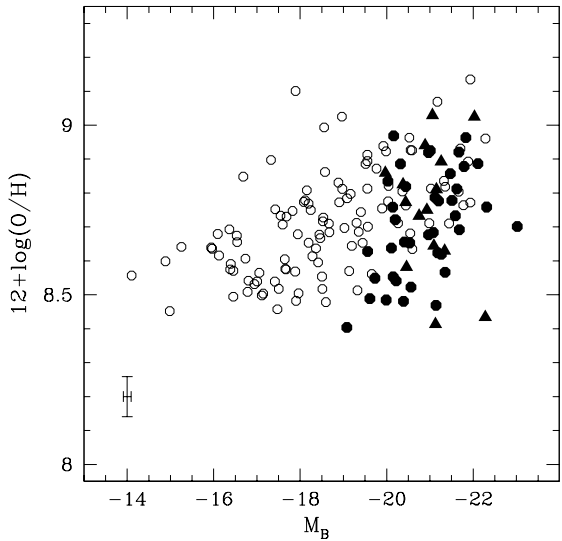}
\caption{Luminosity--metallicity relation for our sample
of intermediate redshift star-forming cluster galaxies 
(filled triangles), compared with a sample of intermediate 
redshift star-forming field galaxies (filled circles), and
local star-forming galaxies from the NFGS sample (open circles).}
\label{oh_mb}
\end{figure}

Fig.~\ref{oh_mb} shows the relationship between galaxy luminosity and oxygen
abundance for our sample of distant star-forming cluster galaxies compared
with the distant field and local galaxy samples. The local sample shows the
well-established luminosity--metallicity relation (e.g., Skillman et al. 1989;
Zaritsky et al. 1995; Richer \& McCall 1995; Melbourne \& Salzer 2002;
Lamareille et al. 2004; Tremonti et al. 2004).  As expected from
Fig.~\ref{fig:distrib}, our distant cluster and field galaxies are distributed
similarly.

\begin{figure*}
\centering
\includegraphics[clip=,width=0.45\textwidth]{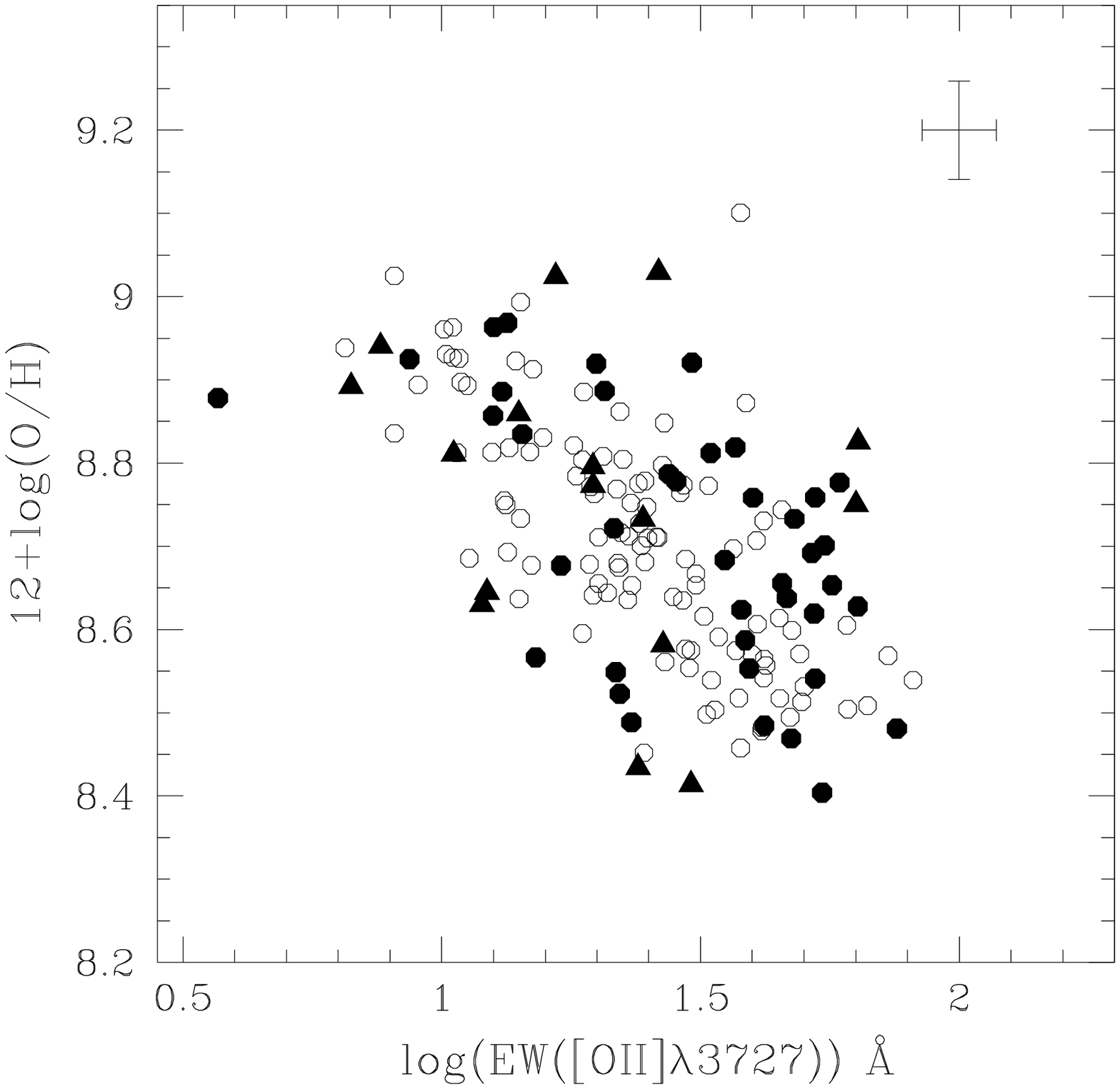}
\includegraphics[clip=,width=0.45\textwidth]{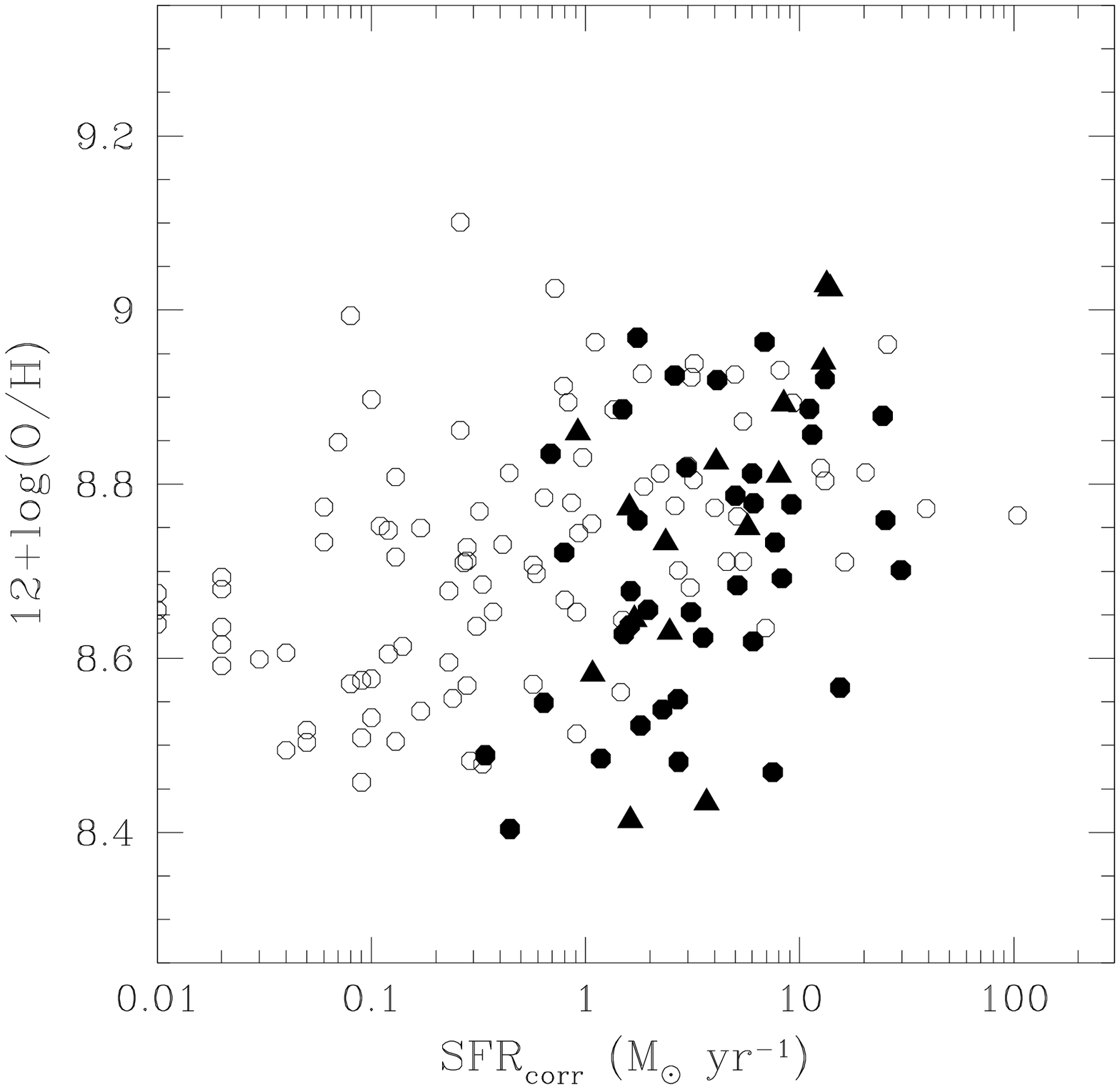}
\caption{Plots showing the relationships between oxygen abundance and 
rest-frame {\oii} equivalent width {\it (left)}, and extinction-corrected 
star formation rate {\it (right)}. Intermediate redshift star-forming
cluster/field galaxies are shown as filled triangles/circles, and the 
NFGS sample of local star-forming galaxies is marked by open circles.}
\label{oh_sfr}
\end{figure*}

The panels of Fig.~\ref{oh_sfr} show the relationship of gas phase 
oxygen abundance versus the rest-frame equivalent width of the {\oii} 
emission line (left) and extinction-corrected star formation rate 
(right). Symbols are the same as in Fig.~\ref{oh_mb}.  
The star formation rates of distant cluster and field galaxies are 
derived using {\hbeta} luminosities, estimated using {\hbeta} equivalent 
widths and $B$-band absolute luminosities following equation~6 of 
Kobulnicky \& Kewley (2004).  The star formation rates are then
calculated following the calibration of Kennicutt (1998).  
The star formation rate of galaxies in the local sample is estimated 
using the extinction-corrected {\halpha} luminosity following the 
calibration of Kennicutt (1998). 

Both distant cluster and field galaxies follow similar relations between
oxygen abundance and {\oii} equivalent width (an indication of star formation
rate per unit luminosity) as the local sample.  However, as seen above, the
intermediate redshift cluster galaxies tend to have lower {\oii} equivalent
widths at a given oxgen abundance than the distant field galaxies.  This is
most noticeable at low oxygen abundances, but does not necessarily imply a
difference in slope, as we may be missing high oxygen abundance cluster
galaxies with very low {\oii} equivalent widths.  As anticipated from
Fig.~\ref{fig:distrib}, there is no evidence for a difference in the
distributions of our intermediate redshift cluster and field galaxies in the
oxygen abundance versus star formation rate diagram.

\section{Discussion and Conclusions}
\label{disc}

We have examined the {\oii}, {\oiii} and {\hbeta} emission line 
equivalent widths, and resultant diagnostic diagrams, oxygen abundances 
and extinction-corrected star formation rates, for a sample of $17$ 
bright ($M_{B} \la -20$), star-forming, mostly disc, cluster galaxies 
at intermediate redshifts, $0.3 \la z \la 0.6$.

The comparison between these distant cluster galaxies and their 
counterparts in the coeval field, from Mouhcine et al.(2006), reveals 
that the properties of the interstellar gas are broadly similar for 
both samples. The primary difference is that the emission line 
equivalent widths of the cluster galaxies are, on average, 
significantly lower than for the field galaxies. However, a fraction 
of the distant cluster galaxies appear to have much higher emission 
line equivalent widths, comparable to the highest seen in the field. 
This tentatively implies a bimodality in the star formation rates per 
unit luminosity of distant cluster galaxies. Our luminous, 
star-forming, intermediate-redshift field galaxies, on the other hand, 
have broad, unimodal distributions, which extend smoothly to ranges 
observed locally only for much fainter galaxies (Mouhcine et al. 2006).

The hint of a bimodality in the star formation rates per unit 
luminosity of distant cluster galaxies, with the majority being 
suppressed, but some apparently enhanced, lends support to a mechanism 
for galaxy evolution in clusters which causes a temporary increase in 
the star formation rate of galaxies prior to a decline in their star 
formation. Indications of this have also been found by considering 
the Tully-Fisher relation of cluster and field samples from which 
the galaxies studied in this paper were drawn (Milvang-Jensen et al. 
2003, Bamford et al. 2005).  In these studies star-forming cluster 
galaxies were found to have, on average, brighter $B$-band 
luminosities than those in the field, at a given rotation velocity.  
This is suggestive of a recent enhancement of star formation for the 
cluster galaxies.

We have examined the oxygen abundances for our galaxies, but are unable 
to discriminate between our cluster and field samples in terms of their 
chemical and ionisation properties. This may be due to a true lack of 
a difference between the two samples, perhaps because the star-forming 
galaxies in distant clusters have recently entered the cluster 
environment from the field. Two galaxies in our cluster sample show 
oxygen abundances significantly lower, but with similar emission line 
equivalent widths, than what is seen locally for similar luminosities. 
One of these two objects has measured rotation velocity and emission 
scale length, i.e., $\vrot \approx {\rm 248\,km\,s^{-1}}$ and 
$\rdspec = {\rm 7.7\,kpc}$. This object is similar to the sub-population 
of massive and large galaxies in the intermediate redshift field that 
are offset with respect to the local luminosity--metallicity relation. 
Given its rotation velocity and physical size, this object is likely to 
evolve into a massive, metal-rich galaxy in the local universe, rather 
than fading into a dwarf galaxy. While this galaxy is only $1 \sigma$ 
away from the cluster redshift, it does lie at a projected distance 
from the cluster centre of $0.8$ times the cluster virial radius, and 
is therefore potentially a recent arrival from the field. 
The other object with low oxygen abundance is at the cluster redshift 
and lies at a projected distance from the cluster centre of $0.3$ 
times the cluster virial radius. It may be a recent addition to the 
cluster, but there is no evidence for or against this. Alternatively, 
the similarities between intermediate redshift field and cluster 
galaxies we have found could be due to the lack of the required 
statistical power to measure the differences, due to our small sample 
size. However, we have established some upper limits on the possible 
differences between the samples.
 
The sample considered in this paper is not large or diverse enough to 
address the issue of differential evolution of cluster and field galaxies 
quantitatively. We have been forced to combine galaxies from a number of 
clusters with a range of redshifts and velocity dispersions, in order to 
produce a usefully sized sample. In doing so we have necessarily 'blurred 
out' any variations which may exist with respect to cluster mass and 
redshift. The galaxy clusters studied here are among the most massive 
candidates at intermediate redshifts ($\sigma_{cl} \approx 750$--$1350$). 
They are also located over a fairly narrow redshift range 
($z = 0.31$--$0.56$). Our results and conclusions are thus restricted to 
massive clusters at $z \sim 0.4$. 

There is as yet little constraints on the form that the variation of 
galaxy chemical evolutionary status takes with environment; is it a 
smooth transition with cluster mass, or are all galaxies in environments 
above a particular threshold density equally affected?
An expectation for the relationship between environment and the chemical 
evolutionary status of its constituent galaxies may be inferred from the 
trends observed for star formation.
The total star formation rate per cluster mass (Finn et al. 2005), 
the fraction of star-forming cluster galaxies, and the star formation
rates of individual cluster galaxies (Poggianti et al. 2006) correlate
strongly with cluster mass, in the sense that clusters with larger 
velocity dispersions tend to have systematically lower star-formation 
activity, and are populated by more passive galaxies (but see also 
Kodama et al. 2004b).
An anti-correlation between the cluster X-ray luminosity and the total 
star formation rate per cluster mass (Homeier et al. 2005) offers 
support for the dependence of the normalized star formation rate on 
cluster mass. The correlation between star formation activity and 
cluster mass seems to be traced by both high and low redshift clusters 
(as suggested by Lewis et al. 2002, Gomez et al. 2003). However, 
Goto (2005) has found, using a sample of low redshift cluster galaxies 
as faint as $M_r=-20.38$, drawn from the Sloan Digital Sky Survey, 
that the fraction of late-type galaxies and the mass-normalized star 
formation rate do not significantly depend on cluster mass.

In order to fully quantify the effect of environment on the chemical 
evolution of galaxies, future studies must thus overcome the degenerate
effects of cluster mass and redshift on galaxy properties. They will 
therefore require considerably larger samples than are considered 
here, spanning a range of cluster masses at a variety of redshifts.

Another missing piece of this puzzle is the behaviour of faint 
galaxies. In our sample we consider only the brightest galaxies, 
with $M_{B} \la -20$.  The effect of environment on the chemical 
evolution of galaxies as a function of galaxy luminosity and mass 
is unknown at intermediate redshifts. Investigating this topic must 
be postponed until intermediate redshift field and cluster galaxy 
samples are available with large quantities of star-forming galaxies 
spanning a range of galaxy luminosity.

\bsp

\label{lastpage}


\begin{thebibliography}{99}

\bibitem[]{} Baldwin J.A., Phillips M.M., Terlevich R., 1981, PASP, 93, 5
\bibitem[]{} Balogh M. L., Schade D., Morris S. L., Yee H. K. C., 
             Carlberg R. G., Ellingson E.,  1998, ApJ, 504, L75
\bibitem[]{} Balogh M. L., Baldry I. K., Nichol R., Miller C. Bower R., 
             Glazebrook K.,  2004, ApJ, 615, L101
\bibitem[]{} Bamford S.~P., Milvang-Jensen B., Arag{\' o}n-Salamanca
             A., Simard L., 2005, MNRAS, 361, 109
\bibitem[]{} Bamford S.~P., Arag{\' o}n-Salamanca A., Milvang-Jensen B.,
             2006, MNRAS, 366, 308
\bibitem[]{} Beers T.~C., Flynn K., Gebhardt K., 1990, AJ, 100, 32
\bibitem[]{} Butcher H., Oemler A., 1978, ApJ, 219, 18
\bibitem[]{} Butcher H., Oemler A., 1984, ApJ, 285, 426
\bibitem[]{} Cardelli J.A., Clayton G.C., Mathis J.S., 1989, ApJ, 329, 33
\bibitem[]{} Couch W.J., Sharples R.M., 1987, MNRAS, 229, 423
\bibitem[]{} Cowie L.L., et al., 1997, ApJ, 481, L9
\bibitem[]{} Davis M., et al., 2003, SPIE, 4834, 161
\bibitem[]{} Dressler A., 1980, ApJ, 236, 351
\bibitem[]{} Dressler A., Gunn J.E., 1992, ApJS, 78, 1
\bibitem[]{} Dressler A., et al., 1997, ApJ, 490, 577
\bibitem[]{} Ellison S. L., Kewley L. J., 2005, in "The Fabulous Destiny 
             of Galaxies; Bridging the Past and Present", in press 
             (astro-ph/0508627)
\bibitem[]{} Finn R. A., et al., 2005, ApJ, 630, 206
\bibitem[]{} Girardi M. \& Mezzetti M., 2001, ApJ, 548, 79
\bibitem[]{} G\'omez P. L., et al., 2003, ApJ, 584, 210
\bibitem[]{} Goto T., 2005, MNRAS, 356, L6
\bibitem[]{} Hoekstra H., Franx M., Kuijken K., van Dokkum P.~G., MNRAS, 
             333, 911
\bibitem[]{} Homeier N. L., et al., 2005, ApJ, 621, 651
\bibitem[]{} Huchra J.P., Davis M., Latham D., Tonry J. 1983, ApJS,
             52, 89
\bibitem[]{} Jansen R.A., Fabricant D., Franx M., Caldwell N., 
             2000, ApJS, 126, 331
\bibitem[]{} Kashikawa N., et al., 2002, PASJ, 54, 819
\bibitem[]{} Kennicutt R.C., Jr. 1998, ARA\&A, 36, 189
\bibitem[]{} Kewley L.J., Dopita M.A., 2002, ApJS, 142, 35
\bibitem[]{} Kewley, L.J., Geller, M., \& Jansen, R.A., 2004, AJ, 127, 2002
\bibitem[]{} Kobulnicky H. A., Kennicutt Jr. R. C., Pizagno J. L., 1999,
             ApJ, 514, 544
\bibitem[]{} Kobulnicky H. A., et al. 2003, ApJ, 599, 1006
\bibitem[]{} Kobulnicky H. A., Phillips A. C., 2003, ApJ, 599, 1031
\bibitem[]{} Kodama T., et al., 2004a, MNRAS, 350, 1005
\bibitem[]{} Kodama T., et al., 2004b, MNRAS, 354, 1103
\bibitem[]{} Kong X., Cheng F.Z., Weiss A., Charlot S., 2002, A\&A,
             396, 503
\bibitem[]{} Lamareille F., Mouhcine M., Contini T., Lewis I.J., 
             Maddox S.J., 2004, MNRAS, 350, 396
\bibitem[]{} Lewis I.J., et al., 2002, MNRAS, 334, 673
\bibitem[]{} Liang Y.C., Hammer F., Flores H., Gruel N., Ass\'emat F.,
             2004, A\&A, 417, 905
\bibitem[]{} Maier C., Meisenheimer K., Hippelein H., 2004, A\&A,
             418, 475
\bibitem[]{} McCall M.L., Rybski P.M., Shields G.A., 1985, ApJS, 57, 1
\bibitem[]{} McGaugh S.~S., 1991, ApJ, 380, 140
\bibitem[]{} Melbourne J., Salzer J. J., 2002, AJ, 123, 2302
\bibitem[]{} Milvang-Jensen B., Arag{\'o}n-Salamanca A. Hau G.~K.~T.,
             J{\o}rgensen I. Hjorth J., 2003, 339L, 1
\bibitem[]{} Morgan W.W., 1961, PNAS, 47, 905
\bibitem[]{} Moss C., Whittle M., 1993, ApJ, 407L, 17  
\bibitem[]{} Moss C., Whittle M., 2000, MNRAS, 317, 667
\bibitem[]{} Mouhcine M., Bamford S.P., Arag{\'o}n-Salamanca A., 
             Nakamura O., 2006, MNRAS submitted
\bibitem[]{} Mouhcine, M., Lewis, I, Jones, B., Lamareille, F., 
             Maddox, S.J., Contini, T., 2005, MNRAS, 362, 1143 
\bibitem[]{} Nakamura, O., Arag{\'o}n-Salamanca, A., Milvang-Jensen, B.,
             Arimoto, N., Ikuta, C., Bamford, S. P., 2006, MNRAS, 366, 144
\bibitem[]{} Osterbrock D.E., 1989, Astrophysics of Gaseous Nebulae and
             Active Galactic Nuclei (Mill Valley: Unvi. Sci.)
\bibitem[]{} Pagel B.E.J., Edmunds M.G., Blackwell D.E., Chum M.S., 
             Smith G., 1979, MNRAS, 189, 95
\bibitem[]{} Poggianti B. M., et al., 1999, ApJ, 518, 576
\bibitem[]{} Poggianti B. M., et al., 2006, ApJ submitted
\bibitem[]{} Postman M., Lubin L. M., Oke J.B., 2001, AJ, 122, 1125
\bibitem[]{} Richer M.G., McCall M.L., 1995, ApJ, 445, 642
\bibitem[]{} Salzer J.J., MacAlpine G.M., Boroson T.A., 1989, ApJS, 
             70, 479
\bibitem[]{} Seifert W.,  et al., 2000, in Iye M., Moorwood, A.~F., eds,
             Proc. SPIE Vol. 4008, Optical and IR Telescope Instrumentation
             and Detectors p.~96
\bibitem[]{} Skillman E.D., Kennicutt R.C., Hodge P.W., 1989, ApJ, 
             347, 875
\bibitem[]{} Smail I., et al., 1997, ApJS, 110, 213
\bibitem[]{} Stasi\'nska G., 1990, A\&AS, 83, 501
\bibitem[]{} Stocke J.~T., et al., 1991, ApJS, 76, 813 
\bibitem[]{} Tremonti C.A., Heckman T.M., Kauffmann G., et al., 2004, ApJ,
             613, 898
\bibitem[]{} Tully R.B., Fisher J.R., 1977, A\&A, 54, 661
\bibitem[]{} Veilleux S., Osterbrock D. E., 1987, ApJS, 63, 295
\bibitem[]{} Welch B.L., 1937, Biometrika, 29, 2057
\bibitem[]{} White S.D.M, et al., 2005, A\&A in press (astro-ph/0508351)
\bibitem[]{} Zaritsky D., Kennicutt R.C., Huchra J.P, 1994, ApJ, 420, 87
\bibitem[]{} Ziegler B.L., B\"ohm A., J\"ager K., Heidt J., M\"ollenhoff C.,   
             2003, ApJ, 598, 87 

\end{thebibliography}
\end{document}